\documentclass[journal=nalefd,manuscript=letter]{achemso}

\usepackage[version=3]{mhchem} % Formula subscripts using \ce{}
\usepackage{textgreek}

\author{Eric W. Martin}
\altaffiliation{These authors contributed equally.}
\author{Jason Horng}
\altaffiliation{These authors contributed equally.}
\author{Hanna G. Ruth}
\author{Eunice Paik}
\author{Michael-Henr Wentzel}
\author{Hui Deng}
\author{Steven T. Cundiff}
\email{cundiff@umich.edu}
\affiliation[University of Michigan]
{Department of Physics, University of Michigan, Ann Arbor, MI 48109-1040, USA}

\title[Encapsulation Narrows Excitonic Homogeneous Linewidth of Exfoliated \ce{MoSe2} Monolayer]
  {Encapsulation Narrows Excitonic Homogeneous Linewidth of Exfoliated \ce{MoSe2} Monolayer}

\abbreviations{MDCS, TMDC, FWM, OBE}
\keywords{Transition metal dichalcogenides, encapsulated monolayer, homogeneous linewidth, radiative lifetime, multidimensional coherent spectroscopy \bigskip}

\begin{document}

%%%%%%%%%%%%%%%%%%%%%%%%%%%%%%%%%%%%%%%%%%%%%%%%%%%%%%%%%%%%%%%%%%%%%
%% The "tocentry" environment can be used to create an entry for the
%% graphical table of contents. It is given here as some journals
%% require that it is printed as part of the abstract page. It will
%% be automatically moved as appropriate.
%%%%%%%%%%%%%%%%%%%%%%%%%%%%%%%%%%%%%%%%%%%%%%%%%%%%%%%%%%%%%%%%%%%%%

%%%%%%%%%%%%%%%%%%%%%%%%%%%%%%%%%%%%%%%%%%%%%%%%%%%%%%%%%%%%%%%%%%%%%
%% The abstract environment will automatically gobble the contents
%% if an abstract is not used by the target journal.
%%%%%%%%%%%%%%%%%%%%%%%%%%%%%%%%%%%%%%%%%%%%%%%%%%%%%%%%%%%%%%%%%%%%%
\begin{abstract}

  The excitonic homogeneous linewidth of an exfoliated monolayer \ce{MoSe2} encapsulated in hexagonal boron nitride is directly measured using multidimensional coherent spectroscopy with micron spatial resolution. The linewidth is 0.26 $\pm$ 0.02 meV, corresponding to a dephasing time $T_2 \approx$ 2.5 ps, which is almost half the narrowest reported values for non-encapsulated \ce{MoSe2} flakes. We attribute the narrowed linewidth to Coulomb screening by the encapsulated material and suppression of non-radiative processes. Through direct measurements of encapsulated and non-encapsulated monolayers, we confirm that encapsulation reduces the sample inhomogeneity. However, linewidths measured using photoluminescence and linear absorption remain dominated by inhomogeneity, and these linewidths are roughly 5 times larger than the homogeneous linewidth in even the highest-quality encapsulated materials. The homogeneous linewidth of non-encapsulated monolayers is very sensitive to temperature cycling, whereas encapsulated samples are not modified by temperature cycling. The nonlinear signal intensity of non-encapsulated monolayers is degraded by high-power optical excitation, whereas encapsulated samples are very resilient to optical excitation with optical powers up to the point of completely bleaching the exciton.
  
\end{abstract}

%%%%%%%%%%%%%%%%%%%%%%%%%%%%%%%%%%%%%%%%%%%%%%%%%%%%%%%%%%%%%%%%%%%%%
%% Start the main part of the manuscript here.
%%%%%%%%%%%%%%%%%%%%%%%%%%%%%%%%%%%%%%%%%%%%%%%%%%%%%%%%%%%%%%%%%%%%%

Monolayer van der Waals crystals are a class of materials with widely varying properties and the potential to transform future electronics and optoelectronics \cite{Radisavljevic2011, Baugher2013, Zhang2012}. These atomically thin layered materials can be stacked into heterostructures with new functionalities \cite{Novoselov2016}. A subset of these materials are the semiconducting monolayer transition metal dichalcogenides (TMDCs), which have a direct band gap that makes their electronic transitions optically accessible and thus useful for optoelectronic applications \cite{Ye2015, Wu2015, Salehzadeh2015, Pospischil2014}. The low dimensionality resulting from confinement to a monolayer also means monolayer TMDCs have very strong many-body interactions that result in $\sim$100x larger binding energy of excitons than more conventional III-V semiconductors, such as gallium arsenide. Excitons thus dominate the optical response of semiconductor TMDCs and remain strongly bound at room temperature. The low dimensionality also means excitations in these materials are very sensitive to the external dielectric environment through both screening and introduction of defects.

Encapsulation of monolayer van der Waals crystals in hexagonal boron nitride (hBN) enhances carrier mobility \cite{Dean2010, Cui2015} and significantly improves the monolayer resistance to photodegradation \cite{Ahn2016}. Most notably hBN encapsulation has been shown to greatly reduce the photoluminescence linewidth of \ce{MoSe2}, \ce{MoS2}, \ce{WSe2}, and \ce{WS2} \cite{Urbaszek2017, Ajayi2017, Wierzbowski2017}.

Narrowing of the photoluminescence linewidth of a TMDC monolayer has been taken as an indicator that encapsulation passivates the monolayer and minimizes inhomogeneity resulting from trapped states and defects. However, narrowing of the photoluminescence linewidth can also result from a change in the radiative linewidth of the exciton, which scales with the substrate index. The substrate can further affect the linewidth through its effect on pure dephasing resulting from interactions of the exciton with photons, phonons, and other collective modes. So while it is impressive that encapsulated TMDC photoluminescence linewidths approach the homogeneous limits measured in similar monolayers on different substrates, linear techniques cannot disentangle the linewidth contributions from inhomogeneous broadening, non-radiative processes, and radiative decay. Here we show that encapsulation narrows both the homogeneous and inhomogeneous linewidths such that the inhomogeneous linewidth still dominates. We discuss how our results imply that hBN-encapsulation of monolayer TMDC samples minimizes defects and static doping that result in both long- and short-range disorder. Along with the static lineshape differences, we measure significant permanent modification of the homogeneous linewidth of non-encapsulated samples resulting from temperature cycling and exposure to weak radiation. In contrast, encapsulated samples are very robust to numerous temperature cycles and high radiation exposure.

To directly measure homogeneous linewidths and distinguish the dephasing and decay processes, it is necessary to use nonlinear spectroscopy techniques. Homogeneous linewidths of exciton resonances in bare monolayer TMDC samples have been measured using four-wave-mixing (FWM) based techniques \cite{Moody2015,Kasprzak2016,Jakubczyk2018}. Multidimensional coherent spectroscopy (MDCS) was employed by Moody \emph{et al.} to measure the homogeneous linewidth of \ce{WSe2} grown by chemical vapor deposition (CVD) \cite{Moody2015} and to identify higher order correlated states in a large exfoliated \ce{MoSe2} flake \cite{Hao2017}. MDCS is useful for its ability to unambiguously separate homogeneous and inhomogeneous broadening of exciton linewidths. However, as is typical of most MDCS techniques, this MDCS implementation uses a non-collinear geometry to isolate the signal from the excitation beams, and it is therefor limited by having a relatively large spot size ($\sim$30 \textmu m). The large spot size requires large samples, and large CVD grown samples are known to show worse quality and mid-gap defects as compared to exfoliated samples \cite{Choi2017,Chen2017}. Jakubczyk \emph{et al.} used three-pulse FWM microspectroscopy to measure exfoliated \ce{MoSe2}. This FWM technique provides similar information to MDCS for lineshapes without inhomogeneous dephasing rates \cite{Spivey2007}. The authors demonstrate tremendous variability of the exciton transition energy (greater than 10 meV) and dephasing time (between 0.5 and 1.5 ps) over the single large exfoliated flake \cite{Kasprzak2016}.

\begin{figure}[htp!]
  \includegraphics[width=0.45\textwidth]{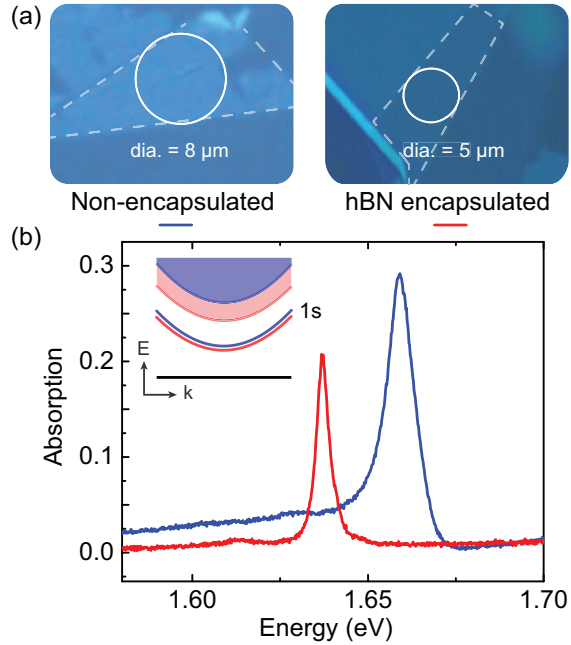}
  \caption{(a) Microscope images of non-encapsulated (left) and hBN encapsulated (right) exfoliated \ce{MoSe2} monolayer samples. These images illustrate the fairly small size of the measured samples. (b) Linear absorption spectra of these two samples, calculated from linear reflectance spectra. The encapsulated sample has a decreased linewidth and decreased intergrated absorption, which are indicators that the sample inhomogeneity has been reduced and non-radiative decay processes have been minimized. (inset) Encapsulation decreases the band gap and exciton binding energy so that the transition energy to the exciton of the encapsulated samples (red line with continuum shaded red) is about 20 meV lower than in the non-encapsulated samples (blue line with continuum shaded blue).}
  \label{fig:linear}
\end{figure}

Here we use MDCS in conjunction with linear reflectance spectroscopy to compare the neutral exciton linewidths of fully hBN-encapsulated and non-encapsulated exfoliated \ce{MoSe2} monolayer samples. Prototypical samples, including heterostructures consisting of exfoliated monolayer materials, are often small. The samples measured here are all between 5 and 8 \textmu m wide. Example images of two of these samples are presented in Fig.~\ref{fig:linear}(a). The small sample size demands the use of collinear techniques for all optical measurements. For linear spectroscopies including photoluminescence and linear absorption, the excitation source is easily distinguished from the sample response. For MDCS it is necessary to distinguish a specific third order response from the excitation sources, linear response, and all other nonlinear responses. The conventional method for isolating the MDCS signal is with k-vector selection \cite{Cundiff2013}, but this method is not congruent with having a tight focus. We have developed a collinear MDCS that has enabled measurements of samples within a diffraction limited spot \cite{Martin2018} based on frequency modulation \cite{Tekavec2007}. For these MDCS measurements we use a spot size of 2 \textmu m.

We compare four high quality samples: two monolayer \ce{MoSe2} mounted on sapphire substrates and two hBN encapuslated monolayer \ce{MoSe2} also mounted on sapphire. The encapsulated monolayers are on a bottom layer of hBN that is approximately 120 nm and a top layer that is approximately 20 nm, measured with atomic force microscopy. We measure encapsulated samples that are of a similar high-quality to others reported in the literature \cite{Wierzbowski2017,Back2018,Scuri2018}, evidenced by the comparable linewidths measured with linear techniques. We plot the linear absorption in Fig.~\ref{fig:linear}(b) in which the exciton of the encapsulated sample has a half width at half maximum (HWHM) linewidth of 2.19 meV. The photoluminescence of this sample has a HWHM linewidth of 1.49 meV. The significant decrease of the total absorption of light by the encapsulated sample is an additional indicator that encapsulation decreases the non-radiative decay processes that contribute to incoherent absorption. We also see that the transition energies of excitons in the two samples differ by about 20 meV. This difference is primarily due to the significant decrease of both the band gap and partially compensating exciton binding energy by encapsulating the monolayer in a high index material \cite{Cadiz2017,Stier2016}. We depict these changes in the inset of Fig.~\ref{fig:linear}(b). With measure the effects of encapsulation on both the inhomogeneous and homogeneous exciton linewidths using MDCS.

\begin{figure}[htp!]
  \includegraphics[width=0.45\textwidth]{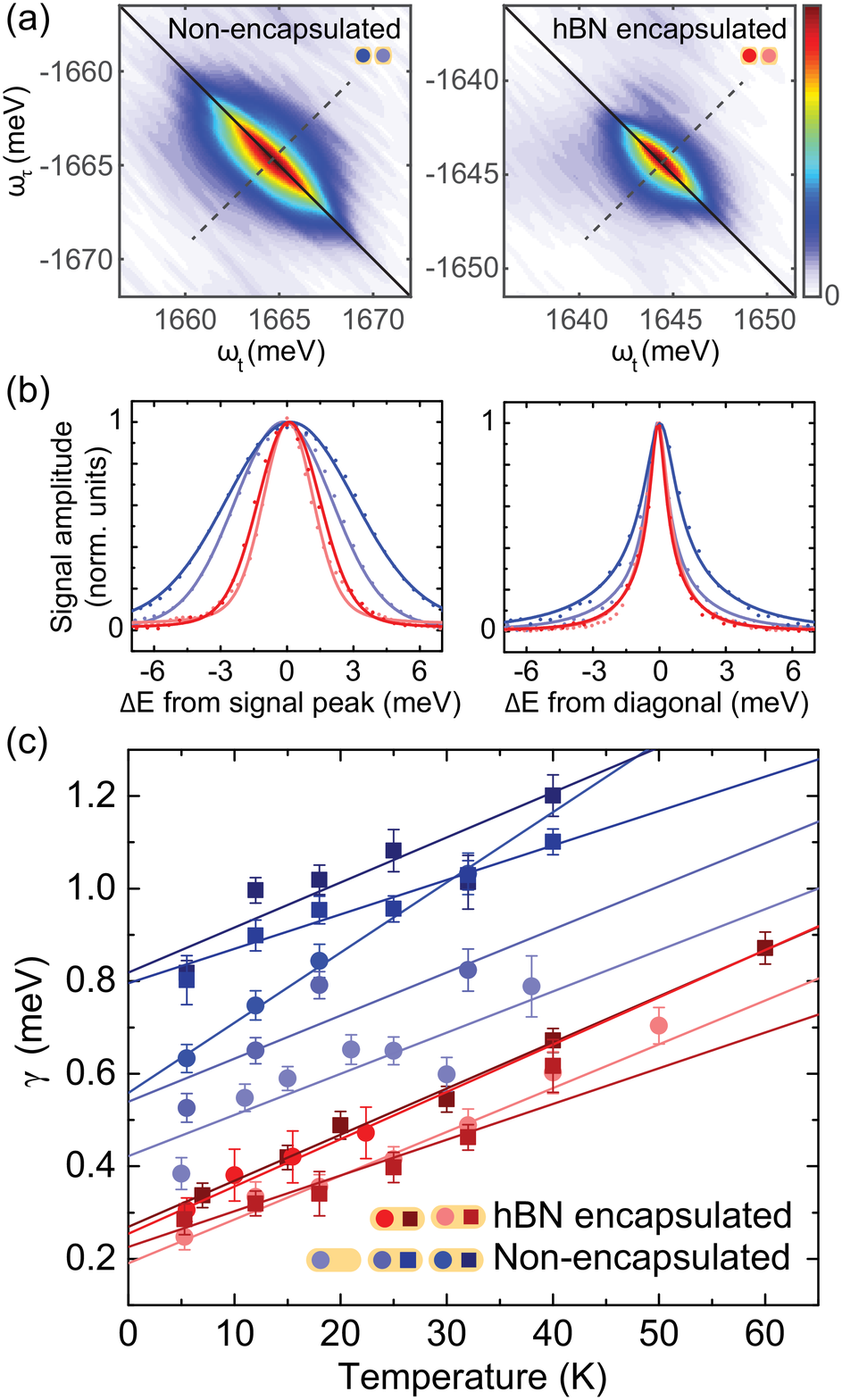}
  \caption{(a) Characteristic low-temperature, low-power multidimensional coherent multidimensional spectra of non-encapsulated \ce{MoSe2} on sapphire and hBN-encapsulated \ce{MoSe2}. In these plots absorption energy ($\omega_{\tau}$) is correlated with emission energy ($\omega_{t}$). (b) Slices along the diagonal (left) of the multidimensional spectrum roughly correspond to the inhomogeneous distribution of exciton resonances. Slices along the cross-diagonal (right) roughly correspond to the homogeneous lineshape. We plot these slices for low temperature, low power measurements of four samples: two non-encapsulated samples in blue and two encapsulated samples in red. Since the diagonal and cross-diagonal slices are correlated, it is essential to fit them simultaneously to determine the homogeneous and inhomogeneous linewidths \cite{Bell2015}. (c) We plot extrapolated zero power linewidths of each sample as a function of temperature. Grouped in the legend, circle data points correspond to a first measurement set of the sample and square data points correspond to a measurement set made after temperature cycling the same sample.}
  \label{fig:MDCS}
\end{figure}

In Fig.~\ref{fig:MDCS}(a) we plot characteristic multidimensional coherent spectra at low temperature using a rephasing pulse sequence. To generate these plots we measure the phase-resolved evolution of an induced nonlinear response as a function of the evolution of a phase-resolved linear response. We measure these responses using a sequence of four pulses in the time domain having relative delays (\texttau~ between the first two pulses, T between the second and third pulse, and t between the last two pulses) that are referenced to a co-propagating continuous-wave laser. Fourier transforming the response with respect to two of the pulse delays yields spectra with two dimensions that correlates absorption ($\omega_{\tau}$) and emission ($\omega_t$) energies of the sample coherences \cite{Cundiff2013, Martin2018}. The evolution of the absorption in this type of MDCS, called a rephasing measurement, has the opposite sign to the emission, so these frequencies are negative. Since a single resonance absorbs and emits at the same energy, a distribution of resonances will all fall along the diagonal where $-\omega_{\tau} = \omega_t$. Thus the lineshape of the diagonal slice plotted on the left in Fig.~\ref{fig:MDCS}(b) roughly corresponds to inhomogenous distributions of exciton resonances in the non-encapsulated (blue) and hBN-encapsulated (red) samples. The lineshape of the cross-diagonal slice plotted on the right roughly corresponds to the homogeneous linewidth of those exciton resonances. For a more detailed description of the MDCS experiment, see the Supporting Information.

Though the cross-diagonal linewidth roughly corresponds to the homogeneous linewidth, the lineshape depends on the inhomogeneous distribution. For low inhomogeneity the cross-diagonal lineshape is a Lorentzian. However, in the limit of high inhomogeneity the lineshape is the square root of a Lorentzian \cite{Siemens2010}. Ignoring inhomogeneous broadening and using the wrong fit function when determining a sample's homogeneous linewidth (or dephasing times in photon echo FWM experiments) can thus significantly skew the measurement, up to a factor of $\sqrt{3}$. For inhomogeneous linewidths that are comparable to the homogeneous linewidth, as is the case in these samples, it is essential to simultaneously fit the codependent diagonal and cross-diagonal slices. Here we fit the entire two-dimensional spectrum using an analytical solution to the optical Bloch equations (OBEs) derived by Bell \emph{et al.} \cite{Bell2015}.

We measure the homogeneous linewidth as a function of beam power and sample temperature. Exciton-exciton interactions are density dependent, and so we measure the density dependence of the linewidth to determine their contribution. We increase the power of all three excitation beams equally, and determine the linewidth scaling as a function of the excitation density of a single beam. We estimate that the linewidth linearly broadens with a slope of $4 \times 10^{-13}$ meV cm$^2$, shown in the Supporting Information. We measure this linear dependence up to an excitation density of 10$^{12}$ cm$^{-2}$. For each sample temperature, measured between 5 and 80 K, we extrapolate the power dependence of the linewidth to zero power and plot that as $\gamma$. Exciton-phonon scattering can be suppressed by lowering the sample temperature to nearly 0 K. At low temperatures the phonon broadening is due to acoustic phonons, which has a linear dependence on temperature. In Fig.~\ref{fig:MDCS}(c) we plot $\gamma$ as a function of temperature for the four different samples. The non-encapsulated samples are indicated in blue. The circle data points correspond to a first set of measurements on a sample, where the linewidths are first measured at 5.3 K and the temperature is increased. The square data points correspond to measurements made after a temperature cycle defined by warming the sample up to room temperature and cooling it back down to again start the measurement set at 5.3 K. It is evident from this data that the linewidth of the non-encapsulated monolayer is very sample dependent, which confirms results by Jakubczyk \emph{et al.} \cite{Kasprzak2016, Jakubczyk2018}. We further find significant broadening of the exciton linewidth of non-encapsulated samples with a single temperature cycle. By measuring many points on the sample, we confirm that the broadening effect is not the result of a positioning error. We rather suggest that the broadening is likely due to deposition of molecules such as water on the sample surface. Whether the change results from surface molecules or substrate strain, we demonstrate that the homogeneous linewidth is a sensitive indicator of a change in the sample environment. The hBN-encapsulated monolayer samples are indicated in red. The sample variance is very small, and there is no measurable broadening due to temperature cycling in these samples. This consistency is evidence that defect scattering is minimal in encapsulated samples. The durability of the monolayer with temperature cycling is an important confirmation that experiments on encapsulated samples will be consistent and reproducible.

From the temperature dependence we measure a linewidth broadening of 0.010 $\pm$ 0.001 meV/K by averaging the hBN encapsulated sample linewidths. This exciton-acoustic phonon interaction is similar to those measured for the non-encapsulated samples confirming previous results \cite{Kasprzak2016}, similar to the interaction in ZnSe (slope = 8 \textmu eV/K) \cite{Wagner1997}, and approximately double the interaction GaAs quantum wells (slope = 5 \textmu eV/K) \cite{Schultheis1986}. The most homogeneous sample we measure has an extrapolated zero temperature and zero power linewidth of $\gamma_0 = 0.20 \pm 0.02$ meV, and the average linewidth for the encapsulated samples is $\gamma_0 = 0.23 \pm 0.03$ meV. This average linewidth corresponds to a dephasing time $T_2 = \hbar/\gamma \approx 2.5$ ps. Since the sample variance of the non-encapsulated samples likely results from surface molecules and substrate effects, the broadest linewidths are dominated by non-radiative decay and pure dephasing processes. Though it is important to note that measurements of the homogeneous linewidth can only identify an upper bound of the radiative linewidth, we identify the lowest measured linewidth for the non-encapsulated samples as the nearest to the radiatively limited linewidth for these samples. For this non-encapsulated sample $\gamma_0 = 0.42 \pm 0.05$ meV, which corresponds to $T_2 \approx 1.6$ ps. This latter value is in agreement with previous determinations of the longest $T_2$ times measured in \ce{MoSe2} using FWM \cite{Kasprzak2016} and time-resolved photoluminescence \cite{Robert2016}. The average linewidth for non-encapsulated samples is $\gamma_0 = 0.6 \pm 0.2$ meV.

We attribute the homogeneous linewidth differences between the hBN encapsulated monolayer and the monolayer directly on sapphire to a combination of factors. The decreased defects and static doping in the encapsulated samples of course minimize the variance. One would expect additional narrowing of the homogeneous linewidth of the encapsulated sample due to the difference in the dielectric environments. The radiative linewidth should scale with the substrate refractive index: $\gamma_{rad} \propto 1/n_{top,~bottom}$ \cite{Selig2016}. For a radiatively-limited homogeneous linewidth, one would thus expect an hBN encapsulated sample to have a $\frac{n_{sapph}+n_{vac}}{2\times n_{hBN}} \sim 0.7$ times narrower linewidth than an equivalent sample on sapphire.

\begin{figure}[htp!]
  \includegraphics[width=0.45\textwidth]{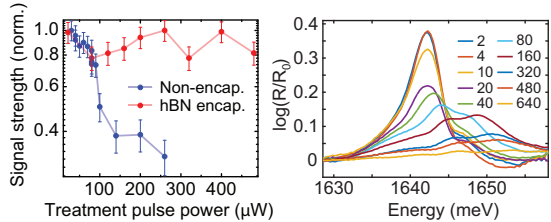}
  \caption{(left) Samples are exposed to a treatment beam of pulses for one minute and turned off. MDCS signal strength is measured as a function of this treatment pulse power, where the measurements are made in ascending order. We find lasting sample damage to non-encapsulated samples plotted in blue, while the encapsulated sample in red is resilient up through powers that saturate the exciton. (right) Single pulse reflectance spectroscopy is used to demonstrate saturation of the exciton in encapsulated samples at high powers that do not damage the nonlinear exciton response.}
  \label{fig:PDep}
\end{figure}

Finally we compare photodegradation of low temperature samples resulting from excitation by resonant pulses. We treat samples by irradiating them with laser light for one minute at the given treatment beam power. The pulsed light is focused to a 2 \textmu m spot and has a repetition frequency of 76 MHz. After each treatment we turn off the treatment beam and measure a multidimensional spectrum with low power, 1 \textmu W/beam, pulses. In Fig.~\ref{fig:PDep} we plot the MDCS signal strength as a function of treatment pulse power. We find that the non-encapsulated samples exhibit lasting damage by beams having powers greater than 45 \textmu W. Measured homogeneous linewidths vary significantly between treatments and scans. The encapsulated samples, however, are resilient up to powers that fully saturate the exciton and have a consistent homogeneous linewidth. We demonstrate saturation of the exciton in an encapsulated sample with single pulse reflectance, similar to an experiment recently presented on non-encapsulated samples \cite{Scuri2018}. We plot reflectance measured with beams having powers between 2 and 640 \textmu W, a range over which the sample is not damaged. Reflectance at the exciton resonance of the encapsulated sample with a beam having a power of 640 \textmu W is saturated. This type of measurement would not be reliable in non-encapsulated \ce{MoSe2}.

In summary, we have measured a significant improvement in sample consistency and stability by encapsulating monolayer \ce{MoSe2} in hBN. In agreement with previous studies, we find that encapsulated samples have narrower inhomogeneous linewidths than non-encapsulated samples. However, we also find that the excitonic homogeneous linewidth is still significantly narrower than the total linewidth. The measurements indicates that inhomogeneity must still be considered in low temperature studies, and the homogeneous linewidth is the upper bound for the radiative linewidth in these samples. Beyond the intrinsic changes to the monolayer by hBN-encapsulation, we demonstrate that encapsulated samples are more robust to high optical excitations. This resilience is essential for lasing applications and ensuring experiments are reproducible.

%%%%%%%%%%%%%%%%%%%%%%%%%%%%%%%%%%%%%%%%%%%%%%%%%%%%%%%%%%%%%%%%%%%%%
%% The "Acknowledgement" section can be given in all manuscript
%% classes.  This should be given within the "acknowledgement"
%% environment, which will make the correct section or running title.
%%%%%%%%%%%%%%%%%%%%%%%%%%%%%%%%%%%%%%%%%%%%%%%%%%%%%%%%%%%%%%%%%%%%%
\begin{acknowledgement}

EWM, HGR, and STC acknowledge the support by the National Science Foundation (NSF) under Grant No. 1622768 and the MCubed Program at the University of Michigan. JH, EP and HD acknowledge the support by the Army Research Office under Award W911NF-17-1-0312 (MURI). MHW acknowledges the support by the University of Michigan Undergraduate Research Opportunity Program.

\end{acknowledgement}

%%%%%%%%%%%%%%%%%%%%%%%%%%%%%%%%%%%%%%%%%%%%%%%%%%%%%%%%%%%%%%%%%%%%%
%% The same is true for Supporting Information, which should use the
%% suppinfo environment.
%%%%%%%%%%%%%%%%%%%%%%%%%%%%%%%%%%%%%%%%%%%%%%%%%%%%%%%%%%%%%%%%%%%%%
\begin{suppinfo}

Detailed schematic and description of the MDCS experiment and
experimental procedures can be found in the Supporting Information:

\end{suppinfo}

%%%%%%%%%%%%%%%%%%%%%%%%%%%%%%%%%%%%%%%%%%%%%%%%%%%%%%%%%%%%%%%%%%%%%
%% The appropriate \bibliography command should be placed here.
%% Notice that the class file automatically sets \bibliographystyle
%% and also names the section correctly.
%%%%%%%%%%%%%%%%%%%%%%%%%%%%%%%%%%%%%%%%%%%%%%%%%%%%%%%%%%%%%%%%%%%%%
\bibliography{TMDC}

\end{document}